\documentclass[aps,pre,twocolumn,showpacs,%
superscriptaddress,preprintnumbers,amssymb]{revtex4-1}
\usepackage{graphicx}
\usepackage[colorlinks=true,linkcolor=blue,%
pagecolor=blue,citecolor=blue,%
urlcolor=blue,anchorcolor=black,%
bookmarksnumbered=true,%
bookmarksopen=true]{hyperref}
\usepackage{dcolumn}
\usepackage{bm}
\renewcommand{\frac}[2]{\displaystyle{#1 \over #2}}
\begin{document}
\title{Dust coupling parameter of 
radio-frequency-discharge complex plasma 
under microgravity conditions}
\author{D.~I.~Zhukhovitskii} \email{dmr@ihed.ras.ru}
\homepage{http://oivtran.ru/dmr/}
\affiliation{Joint Institute of High Temperatures, Russian 
Academy of Sciences, Izhorskaya 13, Bd.~2, 125412 
Moscow, Russia}
\affiliation{Moscow Institute of Physics and Technology, 9 
Institutskiy per., Dolgoprudny, Moscow Region, 141701 
Russia}
\author{V.~N.~Naumkin}
\affiliation{Joint Institute of High Temperatures, Russian 
Academy of Sciences, Izhorskaya 13, Bd.~2, 125412 
Moscow, Russia}
\author{A.~I.~Khusnulgatin}
\affiliation{Moscow Institute of Physics and Technology, 9 
Institutskiy per., Dolgoprudny, Moscow Region, 141701 
Russia}
\affiliation{Joint Institute of High Temperatures, Russian 
Academy of Sciences, Izhorskaya 13, Bd.~2, 125412 
Moscow, Russia}
\author{V.~I.~Molotkov}
\affiliation{Joint Institute of High Temperatures, Russian 
Academy of Sciences, Izhorskaya 13, Bd.~2, 125412 
Moscow, Russia}
\author{A.~M.~Lipaev}
\affiliation{Joint Institute of High Temperatures, Russian 
Academy of Sciences, Izhorskaya 13, Bd.~2, 125412 
Moscow, Russia}
\date{\today}
\begin{abstract}
Oscillation of particles in a dust crystal formed in a 
low-pressure radio-frequency gas discharge under 
microgravity conditions is studied. Analysis of experimental 
data obtained in our previous study shows that the oscillations 
are highly isotropic and nearly homogeneous in the bulk of a 
dust crystal; oscillations of the neighboring particles are 
significantly correlated. We demonstrate that the standard 
deviation of the particle radius-vector along with the local 
particle number density fully define the coupling parameter of 
the particle subsystem. The latter proves to be of the order of 
100, which is two orders of magnitude lower than the 
coupling parameter estimated for the Brownian diffusion of 
particles with the gas temperature. This means significant 
kinetic overheating of particles under stationary conditions. A 
theoretical interpretation of the large amplitude of oscillation 
implies the increase of particle charge fluctuations in the dust 
crystal. The theoretical estimates are based on the ionization 
equation of state for the complex plasma and the equation for 
the plasma perturbation evolution. They are shown to match 
the results of experimental data processing. Estimated order 
of magnitude of the coupling parameter accounts for the 
existence of the solid--liquid phase transition observed for 
similar systems in experiments.
\end{abstract}
\pacs{52.27.Lw, 82.70.-y, 87.15.nt}
\maketitle
\section{\label{s1}INTRODUCTION}

The ionized gas including dust particles typically in the 
range from tens nanometer to thousands of micrometers is 
commonly called the complex (or dusty) plasmas 
\cite{1,2,3,5,6,9}. Such a system makes it possible to study 
fundamental processes in the strong coupling regime on the 
kinetic level through the observation of individual particles. 
In ground-based experiments, gravity has a dominant effect 
on the structures formed in complex plasmas so that 
three-dimensional (3D) dust clouds with an adoptable level 
of the cloud homogeneity cannot be created. In contrast, 
relatively homogeneous dust particle structures are realized 
under microgravity conditions either in parabolic flights 
\cite{10,11,12,13,14} or onboard the International Space 
Station (ISS) \cite{10,15,16,17,18,019,19}. Due to the high 
mobility of electrons, particles acquire a significant 
(macroscopic) negative electric charge, which leads to the 
great Coulomb coupling parameter of the dust subsystem 
$\Gamma$ \cite{1,2,3,4,5,6,8,9}. Thus, such a subsystem 
forms a 3D dust crystal, which, in principle, can undergo 
phase transitions, in particular, the solid--liquid first--order 
transition. In the ground-based experiment, such a transition 
was first observed in study \cite{89}.

Although complex plasmas are open nonequilibrium 
systems, there are grounds to consider a first-order 
transition observed in the dust subsystem similar to that in 
an equilibrium system. A close analog of the dust 
subsystem is a model system known as the one-component 
plasma (OCP) \cite{88}. Then, one can expect the coupling 
parameter of the order of $10^2$ for the solid--liquid 
binodal of the dust subsystem. However, under the 
conditions of PK-3 Plus laboratory onboard the ISS 
\cite{18,92,69}, this parameter amounts to $10^4$ due to a 
significant particle charge. Taking into account the Debye 
screening of the particle charge cannot reduce this 
parameter because the Debye length is on the same order as 
the interparticle distance. Inclusion of the ion-neutral 
collisions into calculation of the particle charge can reduce 
$\Gamma$ at most by an order of magnitude, which is 
insufficient to account for the solid--liquid binodal (and is 
incompatible with the particle oscillation amplitude).

We suggest that the particle kinetic temperature that 
appears in $\Gamma$ is highly increased as compared to 
the gas temperature and the temperature of the particle 
material. Thus, we suggest the anomalous kinetic heating of 
particles in the dust cloud that takes place under stationary 
conditions and does not imply the development of 
instability. Note that the anomalous heating was observed 
in two-dimensional (2D) dust crystal \cite{91} and at the 
bottom of the dust cloud trapped in a striation \cite{90}. In 
both cases, the overheating was a result of the development 
of instability. There is evidence that the kinetic temperature 
of particles in PK-4 experiments is of the order of 
$0.8\;{\mbox{eV}}$
 \cite{50}, which is much higher than the gas temperature 
but it is still insufficient to reduce $\Gamma$ properly.

In this study, we continue to analyze the results obtained in 
the PK-3 Plus experiments \cite{69} and propose the 
method of determination of the particle kinetic temperature. 
Based on the Wigner--Seitz cell model for the dust crystal, 
we show that $\Gamma$ can be only expressed in terms of 
the cell radius and the particle radius-vector standard 
deviation. Thus, the information on the particle charge, 
which cannot be defined for a dust cloud, and on the 
particle velocity, which also cannot be determined due to a 
coarse time resolution of video recording, is unnecessary. 
We arrive at very high kinetic temperature of the particles 
in a stationary dust crystal, which proves to be two orders 
of magnitude higher than that of the gas. We proposed the 
interpretation of the effect of anomalous heating based on 
treatment of the particle charge fluctuations that increase 
significantly the particle oscillation amplitude. On the basis 
of the ionization equation of state (IEOS) \cite{64,69}, 
which is valid both for the stationary and perturbed local 
plasma parameters, and of the equation for propagation of a 
perturbation in the dust crystal, we show that the charge 
fluctuations in the dust crystal have much greater amplitude 
than that for a solitary particle in an infinite plasma. 
Eventually, we obtain the estimation $\Gamma \sim 10^2$ 
and demonstrate that this order of magnitude is compatible 
with the possibility of observation of the solid--liquid phase 
transition.

The paper is organized as follows. In Sec.~\ref{s2}, we 
propose the method of experimental data processing and 
analyze the accuracy of particle coordinate determination. 
In Sec.~\ref{s3}, we discuss main peculiarities of the 
particle oscillation. In Sec.~\ref{s4}, the theory of dust 
charge fluctuations is developed and the relation between 
charge fluctuations and the particle oscillation amplitude is 
revealed. The results of theoretical estimations and of 
experimental data processing are compared and discussed 
in Sec.~\ref{s5}. The results of this study are summarized 
in Sec.~\ref{s6}.

\section{\label{s2}METHOD OF DETERMINATION 
OF THE COUPLING PARAMETER FOR THE DUST 
SUBSYSTEM}

Consider a dust cloud in the low-pressure gas discharge 
under microgravity conditions. We will assume that the 
coupling parameter for the dust subsystem,
\begin{equation}
\Gamma = \frac{{Z_0^2 e^2 }}{{r_d T_d }} \gg 1, 
\label{e1}
\end{equation}
where $Z_0$ is the average stationary dust particle charge 
in units of the electron charge, $e$
 is the elementary electric charge, $r_d = (3/4\pi n_d 
)^{1/3}$ is the Wigner--Seitz radius for the dust particles, 
$n_d$ is the particle number density, $T_d$ is the dust 
particle kinetic temperature, and the Boltzmann constant is 
set to unity. Then the system is indeed a dust crystal, for 
which the Wigner--Seitz cell model is valid. Within the 
framework of this model, a particle with finite $T_d$ 
oscillates in the spherical harmonic potential of its cell. The 
characteristic oscillation frequency is given by the 
expression $\omega _0^2 = Z_0^2 e^2 /Mr_d^3$ \cite{84}, 
where $M = (4\pi /3)\rho _0 a^3$ is the particle mass, $\rho 
_0$ and $a$
 are the density of particle material and the particle radius, 
respectively (spherical particles are assumed). For typical 
experiments with the particles of the diameter $2a = 
2.55\;\mu {\mbox{m}}$
 (see Sec.~\ref{s3}), we have $\omega _0 \simeq 
600\;{\mbox{s}}^{ - 1} $, which exceeds significantly the 
frequency of video exposure (50 frames/s) for the 
high-resolution camera used in the PK-3 Plus setup 
\cite{69}. This means that the particle displacement 
observed in successive frames must be on the same order as 
its fluctuation amplitude and, therefore, it is not small 
enough to determine $T_d$ directly from the particle 
velocity. Moreover, a direct determination of $Z_0$ for the 
dust crystal is also problematic with the available 
diagnostic tools. However, we will show that the coupling 
parameter $\Gamma$ is fully determined by the local 
number density of particles and their standard deviation 
from their equilibrium positions $\delta r$. Provided that 
the Wigner--Seitz cell model is valid, these positions 
coincide with the centers of corresponding cells.

In what follows, we will treat a single particle in the 
Wigner--Seitz cell with the origin of the coordinate system 
in its center. If we denote the distance of a particle from the 
center by $r$
 and its velocity by $v$
 then the average particle potential energy $M\omega _0^2 
\left\langle {r^2 } \right\rangle /2$
 is equal to its average kinetic energy $M\left\langle {v^2 } 
\right\rangle /2$. Here, angular brackets denote time 
averaging. According to the equipartition theorem 
$M\left\langle {v^2 } \right\rangle /2 = 3T_d /2$, which 
yields $T_d = (Z_0^2 e^2 /3r_d^3 )\left\langle {r^2 } 
\right\rangle $. We substitute the latter expression in the 
definition of the coupling parameter to derive
\begin{equation}
\Gamma = \frac{{3r_d^2 }}{{\left\langle {r^2 } 
\right\rangle }} = 3\left( {\frac{{r_d }}{{\delta r}}} 
\right)^2 , \label{e3}
\end{equation}
where $\delta r = \sqrt {\left\langle {r^2 } \right\rangle } $. 
It is worth mentioning that $Z_0$ cancels in the expression 
(\ref{e3}), therefore, $\Gamma$ proves to be independent 
of the particle charge.

We have to keep in mind that $\delta r$
 is a three-dimensional standard deviation while a sequence 
of video frames provides the information on the projection 
of the deviation on the frame plane. Note that the method of 
3D particle coordinate determination developed in 
Ref.~\cite{69} is not efficient for the particle oscillation 
because the scan time is orders of magnitude longer than 
the particle oscillation period $ \propto \omega _0^{ - 1} $. 
To estimate $\delta r$, we assume isotropy of the particle 
oscillations (which is the case for our system; see 
Sec.~\ref{s3}). Then
\begin{equation}
\delta r = \sqrt {\frac{3}{2}\left( {\left\langle {x^2 } 
\right\rangle + \left\langle {z^2 } \right\rangle } \right)} , 
\label{e031}
\end{equation}
where $x$
 and $z$
 are the two-dimensional Cartesian particle coordinates 
($\left\langle x \right\rangle = \left\langle z \right\rangle = 
0$
 is implied). Hence, (\ref{e3}) can be rewritten in the form,
\begin{equation}
\Gamma = \frac{{2r_d^2 }}{{\left\langle {x^2 } 
\right\rangle + \left\langle {z^2 } \right\rangle }}. 
\label{e4}
\end{equation}
Assuming that the dust crystal is stationary, we can 
estimate $\left\langle {x^2 } \right\rangle$ and $\left\langle 
{z^2 } \right\rangle$ from the particle positions observed in 
successive frames. Obviously, the ratio between the 
frequency of video exposure and $\omega _0$ does not 
matter in this case. The Wigner--Seitz radius $r_d$ can be 
calculated using the method of the local particle number 
density determination \cite{69}.

Below, we will discuss determination of the particle 
coordinates in individual frames. Since the expected 
standard deviation is on the same order as the size of a 
camera pixel ($ \sim 10\;\mu {\mbox{m}}$), a satisfactory 
accuracy of such procedure is problematic. A favorable 
circumstance increases the accuracy dramatically. In fact, 
albeit the particle diameter ($ \sim 3\;\mu {\mbox{m}}$) is 
smaller than a pixel, the light scattered from a particle and 
passed through the optical system of a video camera form 
the Gaussian beam with a typical width of 1.5 pixel and, 
therefore, each particle ``illuminates'' more than ten pixels 
by the light with different intensities. Each pixel codes this 
light intensity by brightnesses. If we group together the 
pixels corresponding to given particle then we can define 
the determined particle radius-vector as the weighted sum
\begin{equation}
{\bf{r}} = \frac{{\sum\limits_{k = 1}^N {I_k {\bf{R}}_k 
} }}{{\sum\limits_{k = 1}^N {I_k } }}, \label{e5}
\end{equation}
where the sum runs over all pixels pertaining to the selected 
group, $I_k$ is the brightness of a pixel ($0 \le I_k \le 
255$), ${\bf{R}}_k$ is the radius-vector of its center. In 
(\ref{e5}), we have to ignore the pixels with the brightness 
comparable to the noise brightness ($I_k < 
I_{\mathrm{th}} $), where $I_{\mathrm{th}} = 20$
 for used video camera. It is the information on the pixel 
brightness involved in (\ref{e5}) that increases the 
coordinate determination accuracy. We performed a 
simulation that allows one to estimate this accuracy.

\begin{figure}
\centering \unitlength=0.24pt
\begin{picture}(800,1650)
\put(-140,-20){\includegraphics[width=8.8cm]{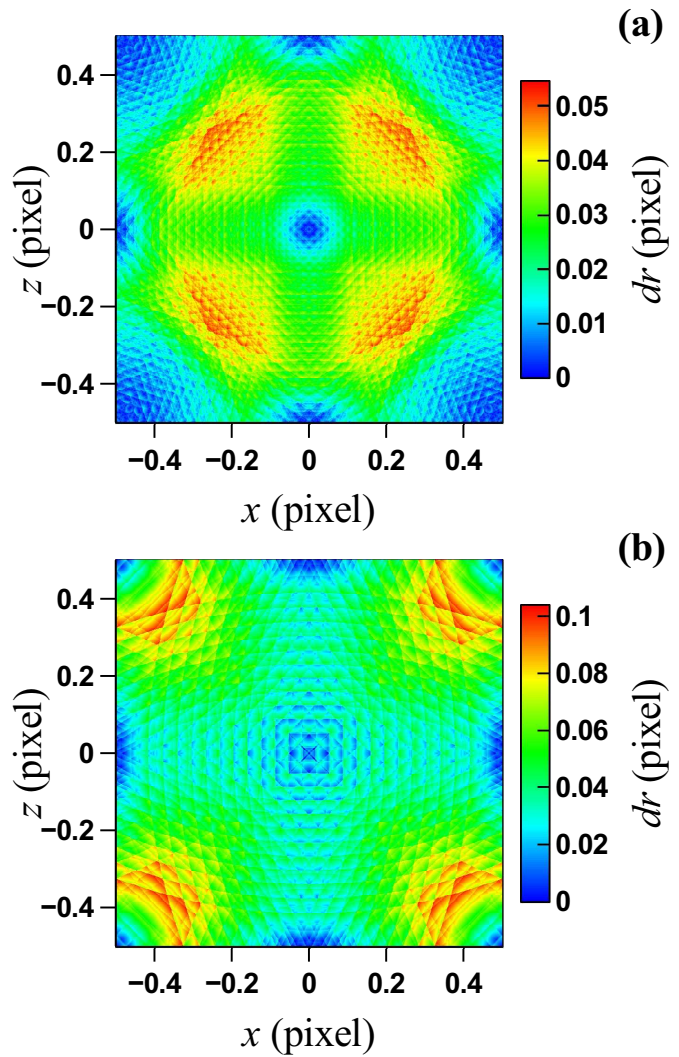}}
\end{picture}
\caption{\label{f1} Distribution of the absolute error 
involved in the proposed method of particle coordinate 
determination over the surface area of a pixel. The distance 
$dr$
 between the center of a true Gaussian beam and the 
determined position of a particle is color-coded. The beam 
amplitude is (a) $A = 100$
 and (b) $A = 50$; for both cases, the width is $w = 
1.5\;{\mbox{px}}$
 and the recognition brightness threshold is 
$I_{\mathrm{th}} = 20$.}
\end{figure}

A Gaussian beam generates a group of pixels with the 
brightness distribution
\begin{equation}
I_k = \left[ {A\exp \left( { - \frac{{\left| {{\bf{R}}_k - 
{\bf{r}}_0 } \right|^2 }}{{w^2 }}} \right)} \right] 
\label{e6}
\end{equation}
for $I_k > I_{\mathrm{th}}$ and $I_k = 0$
 otherwise. In Eq.~(\ref{e6}), $A$
 is the maximum beam intensity, ${\bf{r}}_0$ is the 
radius-vector of the center of a beam that coincides with a 
true particle radius-vector, $w = 1.5\;{\mbox{px}}$
 is a typical beam width, and square brackets denote an 
integral part of a number. The distance between the 
determined and true particle position $dr = \left| {{\bf{r}} - 
{\bf{r}}_0 } \right|$
 is shown in Fig.~\ref{f1} for different ${\bf{r}}_0$ and 
two characteristic maximum pixel brightness. Given 
${\bf{r}}_0 $, $dr$
 was calculated using formulas (\ref{e5}) and (\ref{e6}). 
This quantity defines the absolute accuracy of the particle 
coordinate determination. Obviously, showed distribution is 
periodic in both directions. As is seen, the maximum error 
$dr$
 should be expected if ${\bf{r}}_0$ is varied within a 
single pixel but even in this case, the relative error is below 
10\%. This seems to be an appropriate systematic error as 
compared to the random error, which is noticeably higher 
(Sec.~\ref{s4}). The error decreases with the increase of 
$A$, as it must.
\begin{figure*}
\centering \unitlength=0.24pt
\begin{picture}(800,660)
\put(-575,-20){\includegraphics[width=16.0cm]{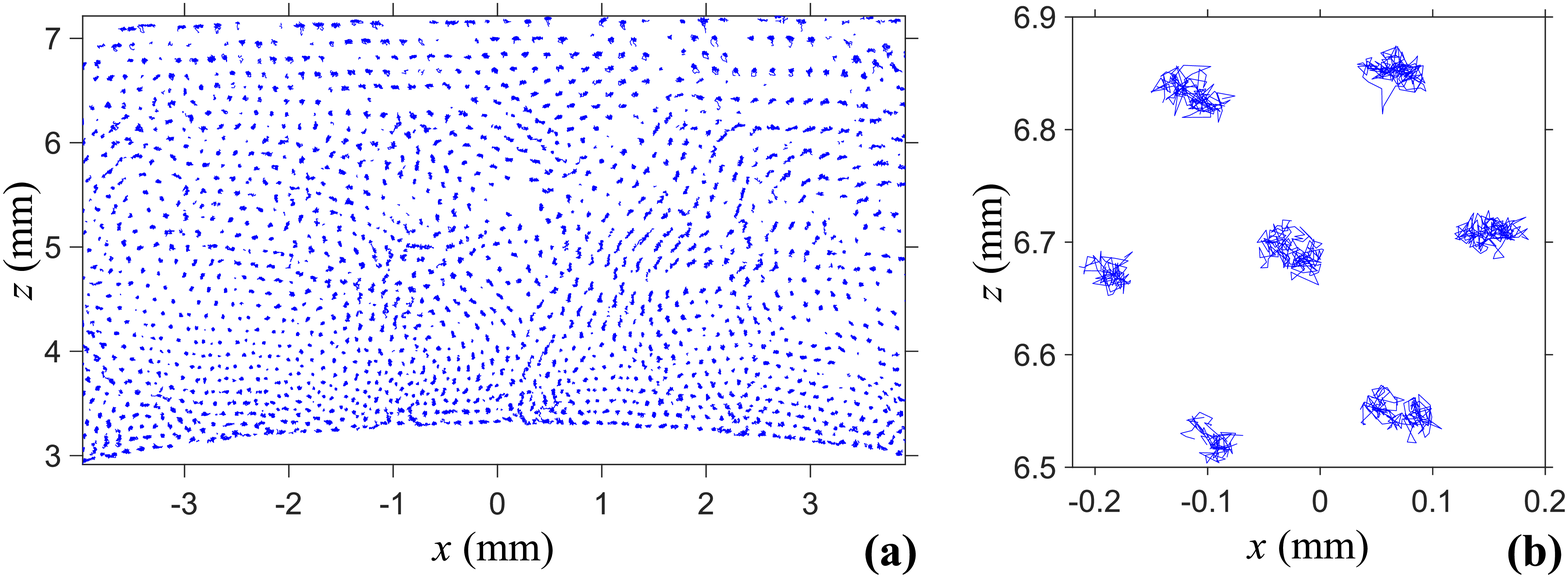}}
\end{picture}
\caption{\label{f2} View of the particle trajectories 
convoluted into clews obtained by superposition of 
successive frames. For different trajectories, the trace 
length varies from 5 to 185 frames. Shown are (a) a view 
from the high-resolution camera and (b) its enlarged 
fragment. The origin of coordinate system coincides with 
the void center, which finds itself almost at the discharge 
chamber axis in the middle of the space between the 
electrodes. The particle diameter is $2.55\;\mu 
{\mbox{m}}$
 and the argon pressure is $10\;{\mbox{Pa}}$.}
\end{figure*}

Application of the above-discussed method of the particle 
coordinate determination allows one to connect the 
positions of individual particles in a sequence of 
consecutive frames by lines and thus to obtain a coarse 
representation of the particle trajectories within their cells 
(Fig.~\ref{f2}). Here and in what follows, we will assume 
that the $Z$-axis is directed toward the upper electrode 
along the symmetry axis of the discharge. The origin of 
axes $XZ$
 is situated in the discharge center. Figure~\ref{f2}(a) 
shows such trajectories for the entire field of view of the 
high-resolution camera. In the bottom of this figure, the 
void boundary is visible and its top corresponds to a 
boundary between the dust crystal and the electrode sheath, 
which is beyond the scope of this treatment. If the number 
of frames for a given particle (particle trace length) is 
sufficiently large, which corresponds to a long observation 
time, then the trajectories assume the form of a clew 
Fig.~\ref{f2}(b). Analysis of the particle positions makes it 
possible to investigate the nature of dust particles 
oscillations.

\section{\label{s3}PROPERTIES OF THE DUST 
PARTICLE OSCILLATIONS}

\begin{figure}
\centering \unitlength=0.24pt
\begin{picture}(800,670)
\put(-130,-20){\includegraphics[width=9.1cm]{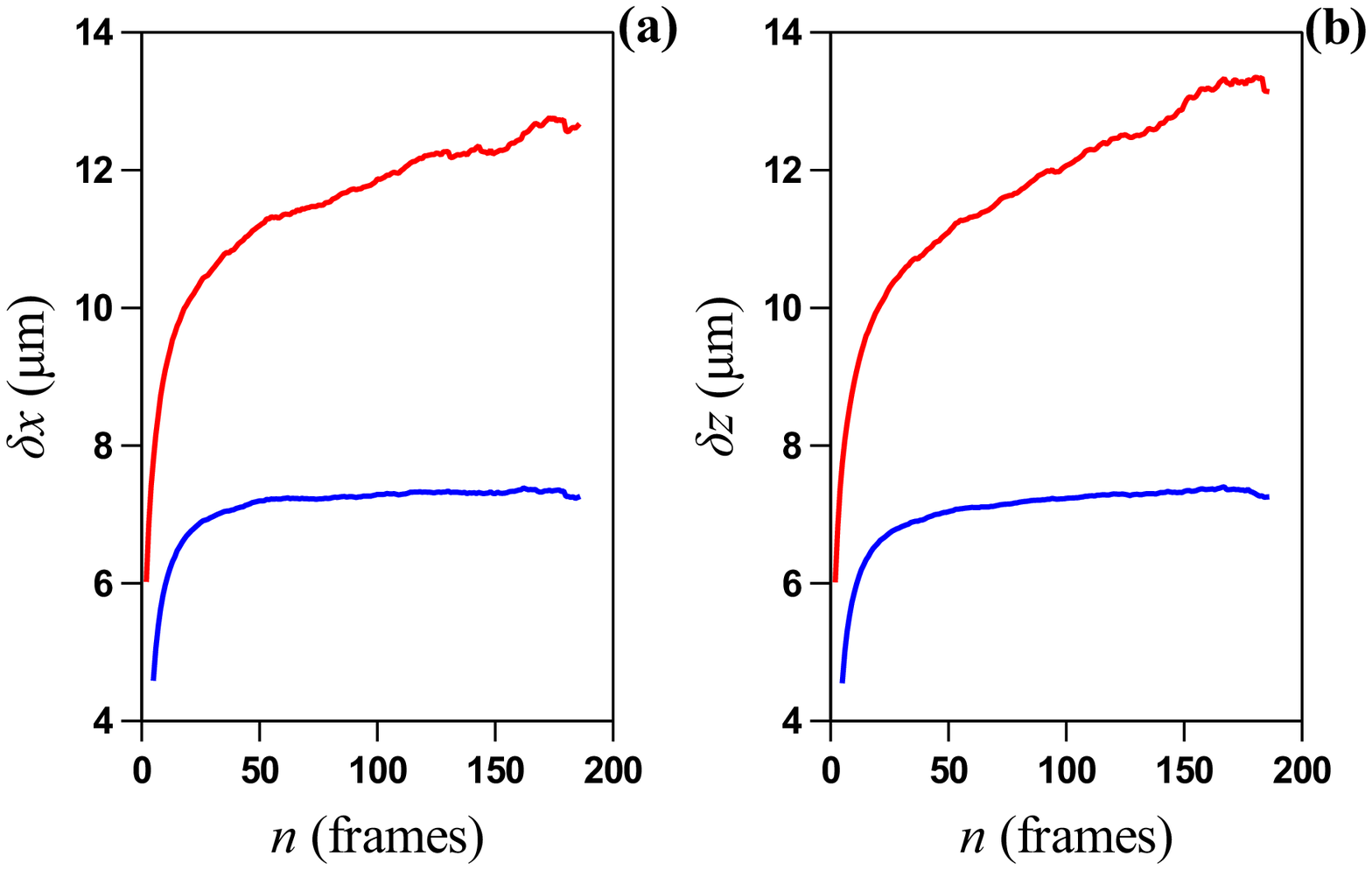}}
\end{picture}
\caption{\label{f3} Standard deviation (a) $\delta x$
 for the coordinate $x$
 and (b) $\delta z$, for $z$
 as a function of the minimum trace length (minimum 
number of successive frames available for the observation 
of an individual particle). Lines show the results of 
averaging over all particle trajectories within the $x$- and 
$z$-corrdinate range shown in Fig.~\ref{f1}(a) (field of 
view of the high-resolution camera); the particle diameter 
and the gas pressure are the same as in this figure. Blue and 
red lines indicate the results obtained with and without the 
trend removal. The particle diameter and the gas pressure 
are the same as in Fig.~\ref{f2}.}
\end{figure}

\begin{figure}
\centering \unitlength=0.24pt
\begin{picture}(800,780)
\put(-140,-20){\includegraphics[width=8.9cm]{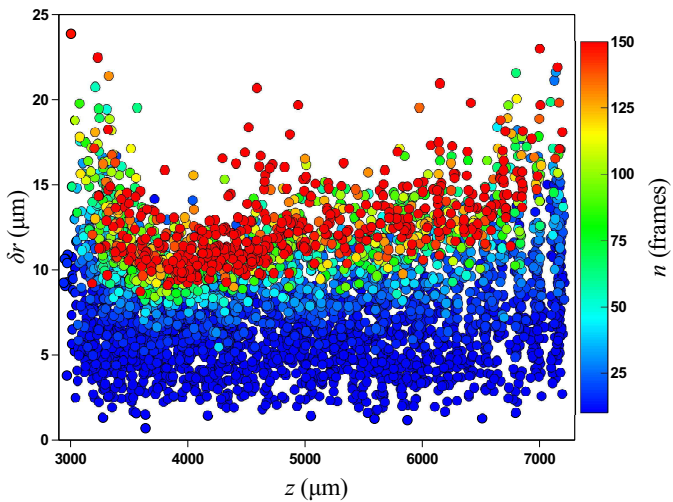}}
\end{picture}
\caption{\label{f4} Three-dimensional standard deviations 
of particles from its equilibrium positions vs the coordinate 
$z$
 for different minimum trace lengths (color-coded). Each 
point indicates averaging over the trace length of an 
individual particle. The range of coordinate $x$
 corresponds to Fig.~\ref{f2}(a), the particle diameter and 
the gas pressure are the same as in this figure. For each 
particle, the trend was removed.}
\end{figure}

In this Section, we will treat a single experiment with the 
particles of the diameter $2.55\;\mu {\mbox{m}}$
 at the gas pressure of $10\;{\mbox{Pa}}$. The first issue is 
the time scale of averaging the particle oscillations. 
Figure~\ref{f3} shows the standard deviations $\delta x$
 and $\delta z$
 for the coordinates $x$
 and $z$, respectively, averaged over all particles whose 
trace length exceeds a definite threshold $n$
 as a function of this threshold. Here, we assume 
homogeneity of a dust crystal and perform averaging over 
all particles in the coordinate range corresponding to 
Fig.~\ref{f2}. One can see that the slope of both curves 
changes significantly at $n \approx 40$
 frames, which corresponds to $0.8\;{\mbox{s}}$, but the 
deviations determined in the quiescent (laboratory) 
coordinate system described above are not stationary even 
after a long time (red curves). This is due to a slow 
large-scale hydrodynamic motion of the dust crystal caused 
by vortices \cite{1,85}, which shift the centers of the 
particle's cells. Thus a trend in the positions of particles is 
formed. Here and in what follows, we will remove this 
trend by polynomial fitting of the particle coordinates (in 
most cases, we used cubic polynomials). With trend 
removal, $\delta x$
 and $\delta z$
 assume stationary values at $n > 40$. As is seen in 
Fig.~\ref{f3}(a) and (b), these values are almost equal, 
which is indicative of the system isotropy. One can 
conclude that the data can be statistically significant if the 
trace length is at least 40 frames. Note that further increase 
of the threshold trace length decreases the sample size. This 
conclusion is justified by Fig.~\ref{f4} that presents 3D 
deviations of individual particles for different minimum 
trace lengths. As is seen, beginning with ca.\ $n = 40$, the 
scatter of $\delta r$
 becomes moderate and it is almost independent of $n$. 
Figure~\ref{f4} also illustrates the dependence $\delta r$
 on the vertical coordinate. A sharp increase of $\delta r$
 at $z < 3500\;\mu {\mbox{m}}$
 is due to the instability that takes place at the void 
boundary that involves the particles in intense motion. At 
$z > 6800\;\mu {\mbox{m}}$, the electrode sheath zone is 
situated. Here, the dust crystal no longer exists, and 
characteristic layered structure is visible. Within these 
limits, $\delta r$
 shows a weak tendency to increase with the increase of 
$z$. This means that the dust crystal is, strictly speaking, 
inhomogeneous with respect to the particle oscillations.
\begin{figure*}
\centering \unitlength=0.24pt
\begin{picture}(800,775)
\put(-580,-20){\includegraphics[width=16.0cm]{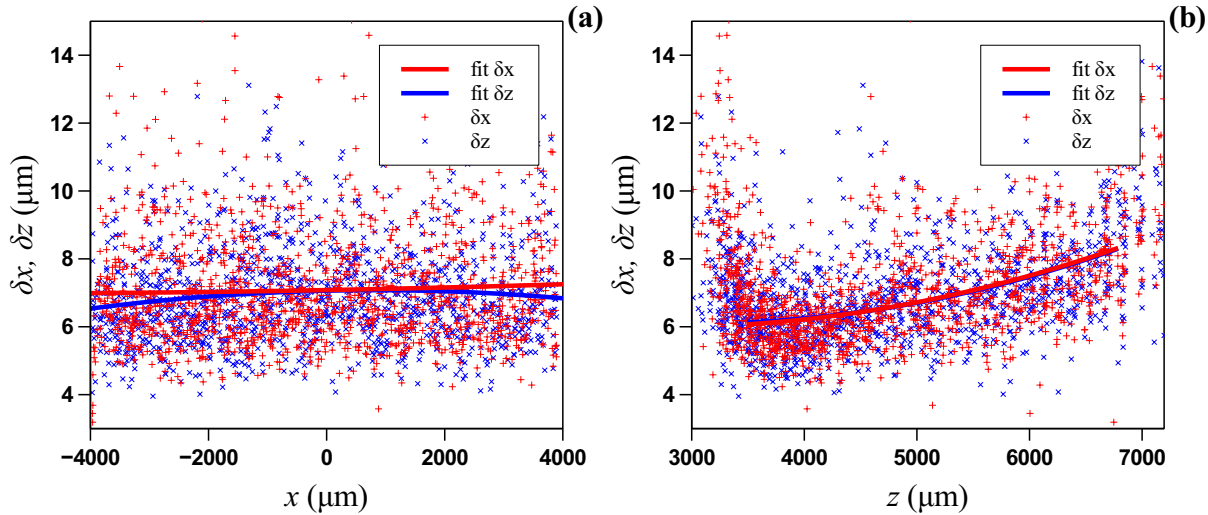}}
\end{picture}
\caption{\label{f5} Standard deviations $\delta x$
 and $\delta z$
 vs (a) the coordinate $x$
 and (b) $z$. Dots show averaging for the individual 
particles, for which the trace length exceeds 40 frames. The 
range of particle coordinates $z$
 in panel (a) is the same as the range of $Z$-axis in panel 
(b) and the range of particle coordinates $x$
 in panel (b) is the same as the range of $X$-axis in panel 
(a). Lines show fitting the experimental results by the 
quadratic polynomials. Red lines and dots indicate $\delta 
x$
 and blue, $\delta z$. The particle diameter and the gas 
pressure are the same as in Fig.~\ref{f2}.}
\end{figure*}

Another illustration of the particle oscillation isotropy and 
weak inhomogeneity is presented by Fig.~\ref{f5}. In this 
figure, the deviations $\delta x$
 and $\delta z$
 are shown only for the individual particles with $n > 40$. 
These data are fitted by the quadratic polynomials. One can 
see that the fitting curves almost coincide, which testifies 
the system isotropy. More exactly, one can state that within 
the accuracy of data processing performed in this study, no 
anisotropy was found. Obviously, the system is 
homogeneous along the $X$-axis and weakly 
inhomogeneous along the $Z$-axis.
\begin{figure}
\centering \unitlength=0.24pt
\begin{picture}(800,570)
\put(-135,-20){\includegraphics[width=9.0cm]{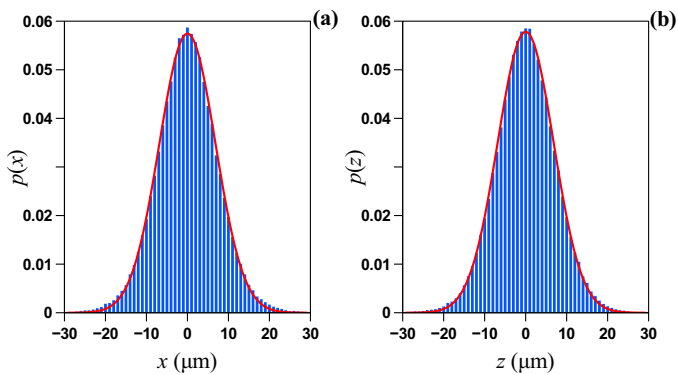}}
\end{picture}
\caption{\label{f6} Probability to find a particle shifted 
from its equilibrium position (a) in the $x$-direction and 
(b) in the $z$-direction. Corresponding shifts are $\Delta x$
 and $\Delta z$, respectively. The minimum trace length is 
40 frames; averaging is performed over all particles in the 
range shown in Fig.~\ref{f2}(a); the particle diameter and 
the gas pressure are the same as in this figure.}
\end{figure}

One more isotropy test is shown in Fig.~\ref{f6}, where the 
probability distribution over the coordinates is presented. 
This distribution was calculated from the coordinates of all 
particles with $n > 40$. It is seen that the distributions over 
both $x$
 and $z$
 are almost the same (which provides another evidence of 
isotropy) and they are well approximated by the Gaussian 
exponent, whence it follows that the particles indeed 
oscillate in the 3D harmonic potential. This justifies the use 
of the Wigner--Seitz cell model. Apparently, slight 
deviation of the distribution from the Gaussian exponent at 
its wings arises from the potential anharmonicity at large 
deviations. It can be the result of some error involved in the 
trend removal procedure.

\begin{figure}
\centering \unitlength=0.24pt
\begin{picture}(800,770)
\put(-7,-20){\includegraphics[width=6.5cm]{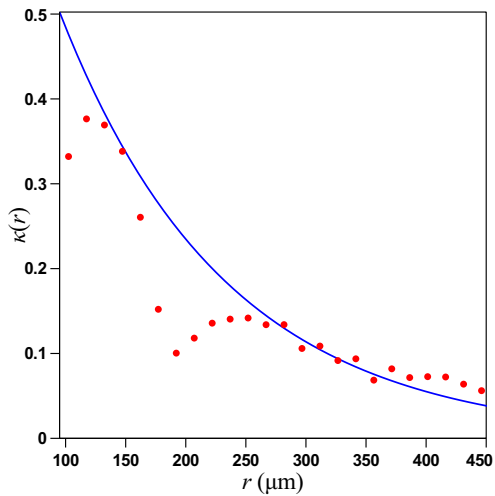}}
\end{picture}
\caption{\label{f7} Correlation coefficient $\kappa$ vs the 
interparticle distance $r$. Dots show averaged experimental 
results. Averaging was performed over all pairs of particles 
at the distance within the intervals of $15\;\mu 
{\mbox{m}}$
 width. Line indicates the curve fit by exponential. The 
minimum trace length is 60 frames, the particle coordinate 
range corresponds to Fig.~\ref{f2}(a), and the particle 
diameter and the gas pressure are the same as in this 
figure.}
\end{figure}

Next, consider the correlation of particle oscillations. We 
failed to estimate the time autocorrelation function for 
individual particles because the decay time for this function 
seems to be $ \lesssim 0.02\;{\mbox{s}}$, i.e., the time 
interval between successive frames is too long, while 
calculation of the pair correlation coefficient is possible. 
We used all pairs of the particles, for which the common 
part $n$
 of the trace lengths is not shorter than 60 frames. Let the 
positions of a pair of particles be defined by the 
radius-vectors ${\bf{r}}_1 = \{ x_1 ,\,y_1 ,\,z_1 \}$ and 
${\bf{r}}_2 = \{ x_2 ,\,y_2 ,\,z_2 \}$ with the origins in 
corresponding centers of the Wigner-Seitz cells. The trend 
is assumed to be removed so that $\left\langle {{\bf{r}}_1 
} \right\rangle = \left\langle {{\bf{r}}_2 } \right\rangle = 
0$. By definition, the pair correlation coefficient is
\begin{equation}
\kappa = \frac{{\sum\limits_{i = 1}^n {{\bf{r}}_{1i} \cdot 
{\bf{r}}_{2i} } }}{{\sqrt {\sum\limits_{i = 1}^n {r_{1i}^2 
} \sum\limits_{i = 1}^n {r_{2i}^2 } } }}, \label{e7}
\end{equation}
where ${\bf{r}}_{1i}$ and ${\bf{r}}_{2i}$ denote the 
particles radius-vectors in the $i$-th frame. In view of a 
weak inhomogeneity of the system in the $z$-direction 
(Fig.~\ref{f5}), one can assume the local (small-scale) 
homogeneity. Due to this and to the global isotropy of the 
system, we have
\begin{equation}
\frac{1}{n}\sum\limits_{i = 1}^n {x_{1i} x_{2i} } \simeq 
\frac{1}{n}\sum\limits_{i = 1}^n {y_{1i} y_{2i} } \simeq 
\frac{1}{n}\sum\limits_{i = 1}^n {z_{1i} z_{2i} } . 
\label{e8}
\end{equation}
Since the radius-vector standard deviations $\delta r_1$ and 
$\delta r_2$ for the particles 1 and 2 can be expressed as
\begin{equation}
\begin{array}{*{20}c}
  {\delta r_{1,2}^2 = \frac{1}{{n - 1}}\left( {\sum\limits_{i 
= 1}^n {x_{1,2i}^2 } + \sum\limits_{i = 1}^n {y_{1,2i}^2 
} + \sum\limits_{i = 1}^n {z_{1,2i}^2 } } \right)} \\
  {} \\
  { \simeq \frac{3}{2}\frac{1}{{n - 1}}\left( 
{\sum\limits_{i = 1}^n {x_{1,2i}^2 } + \sum\limits_{i = 
1}^n {z_{1,2i}^2 } } \right),\quad } \\
\end{array} \label{e9}
\end{equation}
we rewrite Eq.~(\ref{e7}) in the form,
\begin{equation}
\kappa = \frac{3}{2}\frac{{\sum\limits_{i = 1}^n {\left( 
{x_{1i} x_{2i} + z_{1i} z_{2i} } \right)} }}{{(n - 1)\delta 
r_1 \delta r_2 }}. \label{e10}
\end{equation}
The pair correlation coefficient averaged over all 
appropriate pairs of particles is shown in Fig.~\ref{f7} as a 
function of the interparticle distance. One can see that at the 
minimum interparticle distance of $100\;\mu 
{\mbox{m}}$, the correlation is significant. Apparently, 
the maxima at $r = 120$
 and $250\;\mu {\mbox{m}}$
 correspond to the first and the second coordination spheres 
of the dust crystal. If we fit the dependence $\kappa (r)$
 by the exponential $\kappa = e^{ - r/r_c }$ in the range 
$100 < r < 450\;\mu {\mbox{m}}$
 then we obtain the correlation decay length $r_c = 
138\;\mu {\mbox{m}}$, which is close to the average 
interparticle distance (under the conditions of treated 
experiment, it is $n_d^{ - 1/3} = 145\;\mu {\mbox{m}}$). 
At $r > 450\;\mu {\mbox{m}}$, $\kappa (r)$
 decreases to the constant noise level of $0.02$
 and no correlation peaks are observed. Note that at $r = 
450\;\mu {\mbox{m}}$, $\kappa (r)$
 is twice as high as the noise level. It follows from 
Fig.~\ref{f7} that the particle oscillation is correlated with 
that of neighboring particles.

\section{\label{s4}DUST PARTICLE CHARGE 
FLUCTUATIONS AND THE OSCILLATION 
AMPLITUDE}

As was demonstrated in Sec.~\ref{s3}, the particle 
oscillations are highly isotropic. This means that most 
likely, they are not related to the ambipolar electric field 
and the ion flux in complex plasma, neither do they 
originate from development of an instability. We will 
assume that the emergence of oscillations is caused by the 
dust particle charge fluctuations and explore the connection 
between them and the particle oscillations around their 
equilibrium positions. The system is assumed to conform to 
the ionization equation of state (IEOS) for complex plasmas 
with similarity property \cite{22,64,69} based on the fluid 
approach. The fields of particle velocity 
${\bf{v}}(t,\,{\bf{r}})$
 and density $\rho ({\bf{r}}) = Mn_d ({\bf{r}})$
 are solutions of the Euler equation
\begin{equation}
\frac{{\partial {\bf{v}}}}{{\partial t}} + 
({\bf{v}}\cdot\bm{\nabla}){\bf{v}} + \nu {\bf{v}} = 
\frac{1}{\rho }({\bf{f}}_e + {\bf{f}}_{id} - \bm{\nabla}p) 
\label{e100}
\end{equation}
and the continuity equation
\begin{equation}
\frac{{\partial \rho }}{{\partial t}} + \bm{\nabla} \cdot 
(\rho {\bf{v}}) = 0. \label{e200}
\end{equation}
Here, $\nu = (8\sqrt {2\pi } /3)\delta m_n n_n v_{T_n } a^2 
/M$
 is the friction coefficient defining the neutral drag 
\cite{33,6}, $\delta \simeq 1.4$
 is the accommodation coefficient corresponding to the 
diffuse scattering of ions against the particle surface, 
$m_n$ is the mass of a gas molecule, $n_n$ and $v_{T_n } 
= (T_n /m_n )^{1/2}$ are the number density and the 
thermal velocity of gas molecules, respectively, $T_n = 
300\;{\mbox{K}}$
 is their temperature,
\begin{equation}
{\bf{f}}_e = - Zen_d {\bf{E}} = - \frac{{aT_e }}{e}\Phi 
n_d {\bf{E}} \label{e300}
\end{equation}
is the electric field driving force acting on unit volume, 
${\bf{E}} = (T_e /e)\bm{\nabla}\ln n_e$ is the electric 
field strength, $T_e$ is the electron temperature, $n_e$ is 
the electron number density, and $\Phi = Ze^2 /aT_e$ is the 
dimensionless potential of a dust particle;
\begin{equation}
{\bf{f}}_{id} = \frac{3}{8}\left( {\frac{{4\pi }}{3}} 
\right)^{1/3} n_d^{1/3} n_i \lambda e{\bf{E}} 
\label{e400}
\end{equation}
is the ion drag force acting on unit volume, $n_i$ is the ion 
number density, $\lambda$ is the ion mean free path with 
respect to the collisions against gas atoms, and
\begin{equation}
p = \frac{1}{{8\pi }}\left( {\frac{{aT_e }}{{e\lambda ^2 
}}} \right)^2 p^* ,\quad p^* = \Phi ^2 n_d^{*4/3} 
\label{e500}
\end{equation}
is the dust pressure \cite{22}, where $n_d^* = (4\pi 
/3)\lambda ^3 n_d$ is the dimensionless particle number 
density. For a stationary dust crystal, the force balance 
equation yields \cite{22,64}
\begin{equation}
\left( {\frac{{9\pi }}{{128}}} \right)^{1/3} \frac{{n_i 
\lambda }}{{n_d^{2/3} }} = \frac{{aT_e }}{{e^2 }}\Phi . 
\label{e800}
\end{equation}
Equation~(\ref{e800}) is completed by the equation 
defining the particle potential that follows from the OML 
model \cite{54,55}
\begin{equation}
n_e = n_i \theta \Phi e^\Phi , \label{e900}
\end{equation}
where $\theta = \sqrt {T_e m_e /T_i m_i } $, $T_i$ and 
$m_i$ are the ion temperature and mass, respectively, and 
$m_e$ is the electron mass.

The equation for small perturbations of a dust crystal can be 
derived from Eqs.~(\ref{e100}) and (\ref{e200}) linearized 
with respect to small variations of ${\bf{v}}$
 and $\rho $,
\begin{equation}
\frac{{\partial ^2 \psi }}{{\partial t^2 }} + \nu 
\frac{{\partial \psi }}{{\partial t}} = c_s^2 \Delta \psi , 
\label{e700}
\end{equation}
where $\bm{\nabla}\psi = {\bf{v}}$
 and $c_s^2 = dp/d\rho $, where $c_s$ is the velocity of 
dust acoustic waves (DAWs). As was demonstrated in 
Ref.~\cite{64}, $c_s$ is almost independent of the dust 
density distribution inside the dust crystal and can be 
treated as a constant
\begin{equation}
c_s = \frac{{aT_e c_s^* }}{{e\sqrt {6M\lambda } }},\quad 
c_s^{*2} = \frac{{512}}{{27}}\frac{{\Phi _s^2 (\Phi _s + 
1)}}{{(3\Phi _s + 4)(\Phi _s + 2)}}, \label{e801}
\end{equation}
where $\Phi _s$ is the root of the equation
\begin{equation}
\frac{{2\theta e^{\Phi _s } (\Phi _s + 1)}}{{1 - \theta \Phi 
_s e^{\Phi _s } }} = 1 + \frac{2}{{\Phi _s }}. \label{e802}
\end{equation}

Constancy of the DAWs velocity $dp^* /dn_d^* = 
c_s^{*2}$ makes it possible to conclude that the relation 
between the dimensionless ion number density $n_i^* = 
(e^2 \lambda ^3 /aT_e )n_i$ and $n_d^*$ \cite{69} is
\begin{equation}
\Phi = c_s^* \frac{{\sqrt {n_d^* - n_0^* } }}{{n_d^{*2/3} 
}}. \label{e11}
\end{equation}
The relation between $n_i^*$ and $n_d^*$ follows from 
(\ref{e800}) and (\ref{e11}):
\begin{equation}
n_i^* = \frac{2}{\pi }c_s^* \sqrt {n_d^* - n_0^* } . 
\label{e12}
\end{equation}
Here, $n_0^*$ is defined by the ``dust invariant'' $\kappa$ 
\cite{22,69}
\begin{equation}
n_0^* = n_f^* \left( {1 - \frac{{\Phi _f^2 n_f^{*1/3} 
}}{{c_s^{*2} }}} \right),\quad n_f^{* - 1} = (a\kappa T_e 
)^{3/2} \lambda ^3 . \label{e13}
\end{equation}
Equations (\ref{e11}) and (\ref{e12}) define the IEOS for a 
dust crystal. It was demonstrated in \cite{64} that observed 
isotropy of DAWs in an anisotropic dust crystal has an 
important consequence. Namely, the variations of all 
quantities in the perturbation are related by IEOS.

Now turn to the estimation of the charge fluctuations of a 
dust particle. The equation for particle charge kinetics 
follows from the balance of the ion and electron fluxes to 
the particle surface. In the linearized form, this equation 
reads \cite{1,87}
\begin{equation}
\frac{{d\delta Z}}{{dt}} + \nu _f \delta Z = \frac{{2Z\nu 
_f }}{{1 + \Phi _0 }}\theta (t), \label{e14}
\end{equation}
where $Z = Z_0 + \delta Z$
 is the particle charge, $\left\langle Z \right\rangle = Z_0 $, 
and $\left\langle {\delta Z} \right\rangle = 0$
 is the charge fluctuation, which is assumed to be small, 
$\left| {\delta Z/Z_0 } \right| \ll 1$, $\nu _f = av_i (1 + \Phi 
_0 )/4r_{Di}^2$ is the charge relaxation frequency, $v_i = 
\sqrt {8T_i /\pi m_i }$ is the ion thermal velocity, $\Phi _0 
= Z_0 e^2 /aT_e $, $r_{Di} = \sqrt {T_i /4\pi n_i e^2 }$ is 
the ion Debye length, $\theta (t)$
 is a random function that satisfies the conditions
\begin{equation}
\begin{array}{*{20}c}
  {\mathop {\lim }\limits_{t \to \infty } \left[ {t^{ - 1} 
\int\limits_0^t {\theta (t)\,dt} } \right] = 0,} \\
  {} \\
  {\mathop {\lim }\limits_{t \to \infty } \left[ {t^{ - 1} 
\int\limits_0^t {\theta ^2 (t)\,dt} } \right] = 1.} \\
\end{array} \label{e15}
\end{equation}
The left-hand side of Eq.~(\ref{e14}) is responsible for the 
charge relaxation while the right-hand side is the source of 
fluctuations due to discreteness of the ion and electron 
charges.

Equation~(\ref{e14}) is valid for a solitary particle in 
infinite plasma, where $n_i$ is independent of the charge 
fluctuations. In contrast, in the dust crystal, the charge 
fluctuation gives rise to the variation of $n_i$ and $n_e$ at 
the length scale of the order of $r_d$ around the particle, 
and it is these quantities that define the particle charging. 
However, the fluctuation source does not change because it 
is defined by the averages of $n_i$ and $n_e $. At the same 
time, the relaxation frequency $\nu _f$ can change 
dramatically.

In the following, we will derive the equation for the particle 
charge relaxation neglecting the fluctuation source. If we 
treat the charge fluctuation as a plasma perturbation then 
the general equation for its evolution is (\ref{e700}). 
Although it is based on the fluid approach, it can yield 
reasonable results even at the length scales of several 
interparticle distances. Furthermore, we will use this 
equation for a single cell to derive, at least, an 
order-of-magnitude estimate for the amplitude of the charge 
fluctuation. The first term on the left-hand-side of 
(\ref{e700}) corresponds to the perturbation relaxation due 
to the DAWs propagation and the second term, to the 
diffusive relaxation in the overdamped regime. The 
corresponding time scales are $\tau _s = r_d /c_s$ and $\tau 
_{\mathrm{dif}} = \nu r_d^2 /c_s^2$ \cite{86}, 
respectively. Under the conditions of experiment treated in 
Sec.~\ref{s3}, $\tau _{\mathrm{dif}} /\tau _s = \nu r_d /c_s 
= 0.15$, therefore, the diffusive relaxation dominates. 
Accordingly, Eq.~(\ref{e700}) is reduced to
\begin{equation}
\frac{{\partial \psi }}{{\partial t}} = \frac{{c_s^2 }}{\nu 
}\Delta \psi . \label{e16}
\end{equation}
Equation~(\ref{e16}) has a solution $\psi (t,\,{\bf{r}}) = 
\chi (t)\varphi ({\bf{k}} \cdot {\bf{r}})$, which decays 
exponentially in time, $\left| {\bf{k}} \right| = r_d^{ - 1} $, 
and the function $\varphi$ satisfies the condition $(c_s^2 
/\nu )\Delta \varphi = - \varphi /\tau _{\mathrm{dif}} $. 
Then $\chi (t)$
 is a solution of the equation
\begin{equation}
\frac{{d\chi }}{{dt}} + \frac{\chi }{{\tau _{\mathrm{dif}} 
}} = 0. \label{e17}
\end{equation}
It follows from the relation between the particle pressure 
perturbation $\delta p$
 and $\psi$ \cite{64} that $\delta p \simeq - M\nu n_d \psi$ 
in the overdamped regime. Hence, $\delta p$
 also satisfies (\ref{e17}). In view of (\ref{e500}) and 
(\ref{e11}), $\delta Z \propto \delta \Phi$ follows the same 
relaxation rule (\ref{e17}): $d\delta Z/dt + \delta Z/\tau 
_{\mathrm{dif}} = 0$. We compare this with the left-hand 
side of (\ref{e14}) to deduce that for the dust crystal, the 
relaxation frequency changes from $\nu _f$ to $1/\tau 
_{\mathrm{dif}} $. Thus, one can assume that during the 
time $\tau _{\mathrm{dif}} $, the particle motion in the 
charge space is the Brownian diffusion and the drift motion 
can be neglected. This diffusion defines the fluctuation 
amplitude.

Inclusion of the fluctuation source, which is exactly the 
same as in (\ref{e14}), in the equation for charge evolution, 
leads to
\begin{equation}
\frac{{d\delta Z}}{{dt}} + \frac{{\delta Z}}{{\tau 
_{\mathrm{dif}} }} = \frac{{2Z_0 \nu _f }}{{1 + \Phi _0 
}}\theta (t). \label{e18}
\end{equation}
Equations (\ref{e14}) and (\ref{e18}) differ only by the 
relaxation frequency. We apply the fluctuation-dissipation 
theorem to (\ref{e18}) to derive the standard deviation of 
charge fluctuations $\sigma _Z Z_0 = \sqrt {\left\langle 
{\delta Z^2 } \right\rangle } $, where
\begin{equation}
\sigma _Z^2 = \frac{{\nu _f \nu r_d^2 }}{{(1 + \Phi _0 
)Z_0 c_s^2 }}. \label{e19}
\end{equation}
We use (\ref{e11}) and (\ref{e12}) to represent (\ref{e19}) 
in the form,
\begin{equation}
\sigma _Z^2 = \frac{{2a\nu v_i e^2 }}{{\lambda T_i c_s^2 
}}. \label{e20}
\end{equation}
As is seen from (\ref{e20}), $\sigma _Z$ is independent of 
the coordinate. Under typical experimental conditions, it is 
much higher than that for a solitary particle in infinite 
plasma $\sigma _Z^{ - 2} = (1 + \Phi _0 )Z_0$ \cite{1}.

According to the IEOS (\ref{e11}) and (\ref{e12}), the 
charge fluctuations are related to the standard deviation of 
the cell radius $\sigma _d r_d = \sqrt {\left\langle {\delta 
r_d^2 } \right\rangle } $. Since it follows from (\ref{e11}) 
that $\Phi n_d^{*1/6} \simeq c^*$ at $n_d^* \gg n_0^* $, 
we obtain $\sigma _d = 2\sigma _Z$ with due regard for the 
relation $\Phi = Ze^2 /aT_e $. We recall that according to 
Fig.~\ref{f7} the positions of neighboring particles are 
correlated. If we assume that the change of the cell radius 
$r_d$ shifts neighboring particles at $\sqrt {\left\langle 
{\delta r_d^2 } \right\rangle } /3$
 then we can estimate the standard deviation of the particle 
radius-vector $\delta r$
 as
\begin{equation}
(\delta r)^2 = \frac{8}{9}\frac{{a\nu v_i e^2 r_d^2 
}}{{\lambda T_i c_s^2 }}. \label{e21}
\end{equation}
From (\ref{e3}) and (\ref{e21}), the coupling parameter 
can be estimated as
\begin{equation}
\Gamma = \frac{{27}}{8}\frac{{\lambda T_i c_s^2 
}}{{a\nu v_i e^2 }}. \label{e22}
\end{equation}
It follows from (\ref{e21}) and (\ref{e22}) that the relative 
standard deviation $\delta r/r_d$ and $\Gamma$ are 
independent of the coordinate.

\section{\label{s5}RESULTS AND DISCUSSION}

\begin{figure}
\centering \unitlength=0.24pt
\begin{picture}(800,770)
\put(-200,-20){\includegraphics[width=10.6cm]{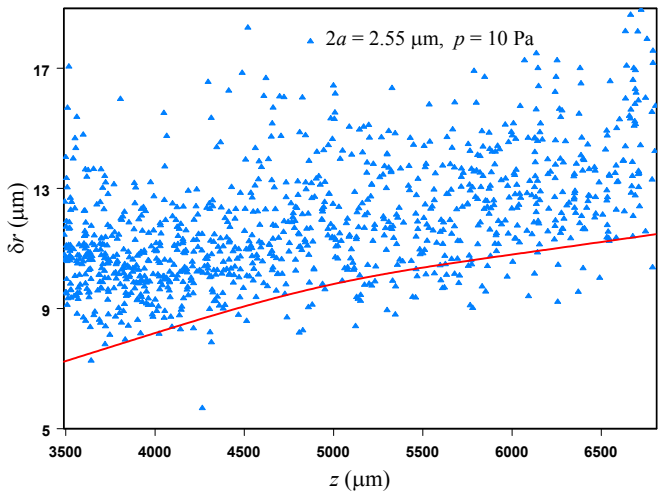}}
\end{picture}
\caption{\label{f8} Three-dimensional standard deviation 
of a particle from its equilibrium position vs the coordinate 
$z$. Dots represent the results of experimental data 
processing and line shows the calculations using formula 
(\ref{e031}). The gas pressure, the particle diameter, and 
the coordinate range are the same as in Fig.~\ref{f2}.}
\end{figure}

\begin{figure}
\centering \unitlength=0.24pt
\begin{picture}(800,770)
\put(-200,-20){\includegraphics[width=10.6cm]{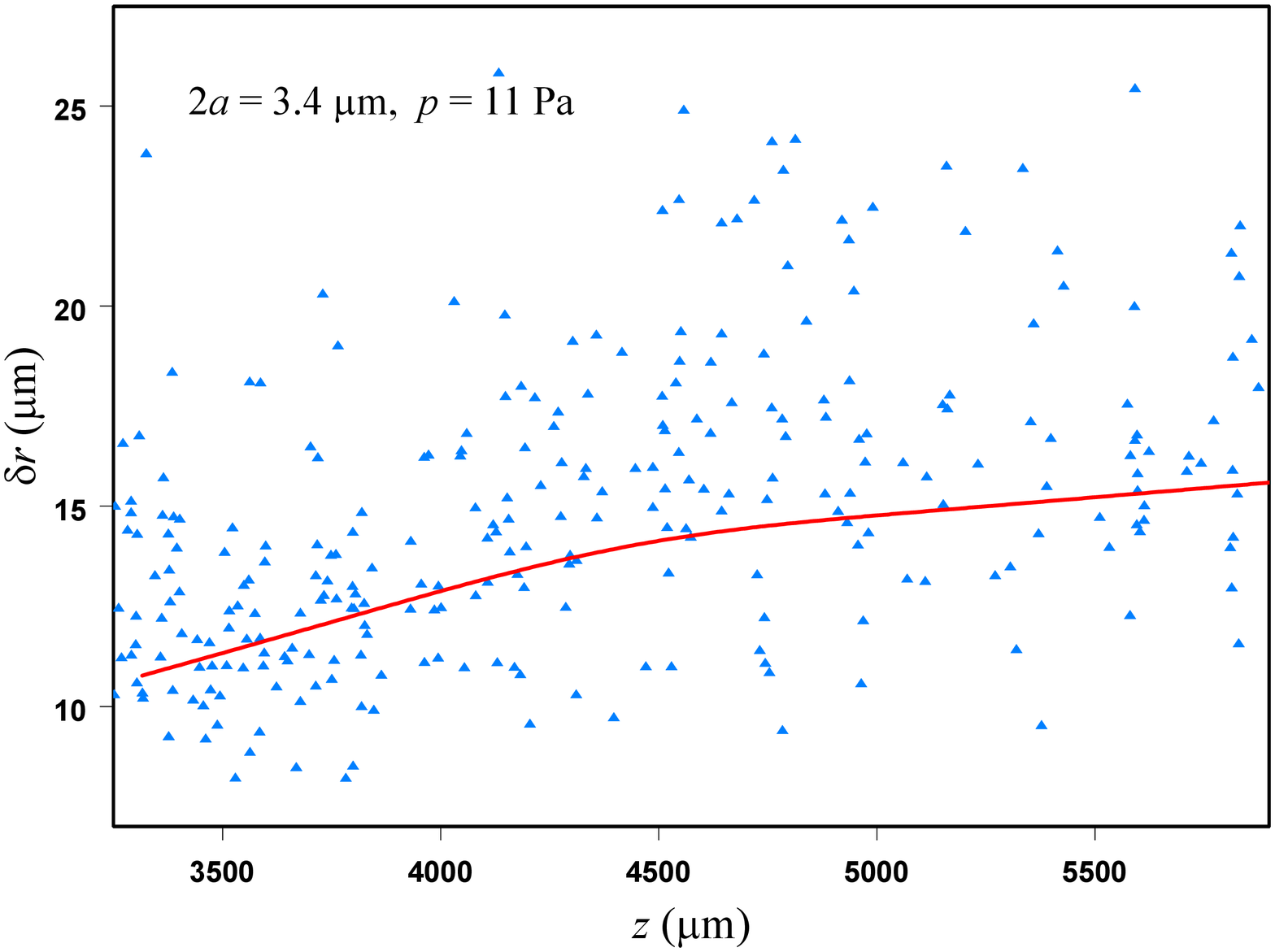}}
\end{picture}
\caption{\label{f9} Same as Fig.~\ref{f8}. The particle 
diameter is $3.4\;\mu {\mbox{m}}$
 and the argon pressure is $11\;{\mbox{Pa}}$.}
\end{figure}

\begin{figure}
\centering \unitlength=0.24pt
\begin{picture}(800,770)
\put(-200,-20){\includegraphics[width=10.6cm]{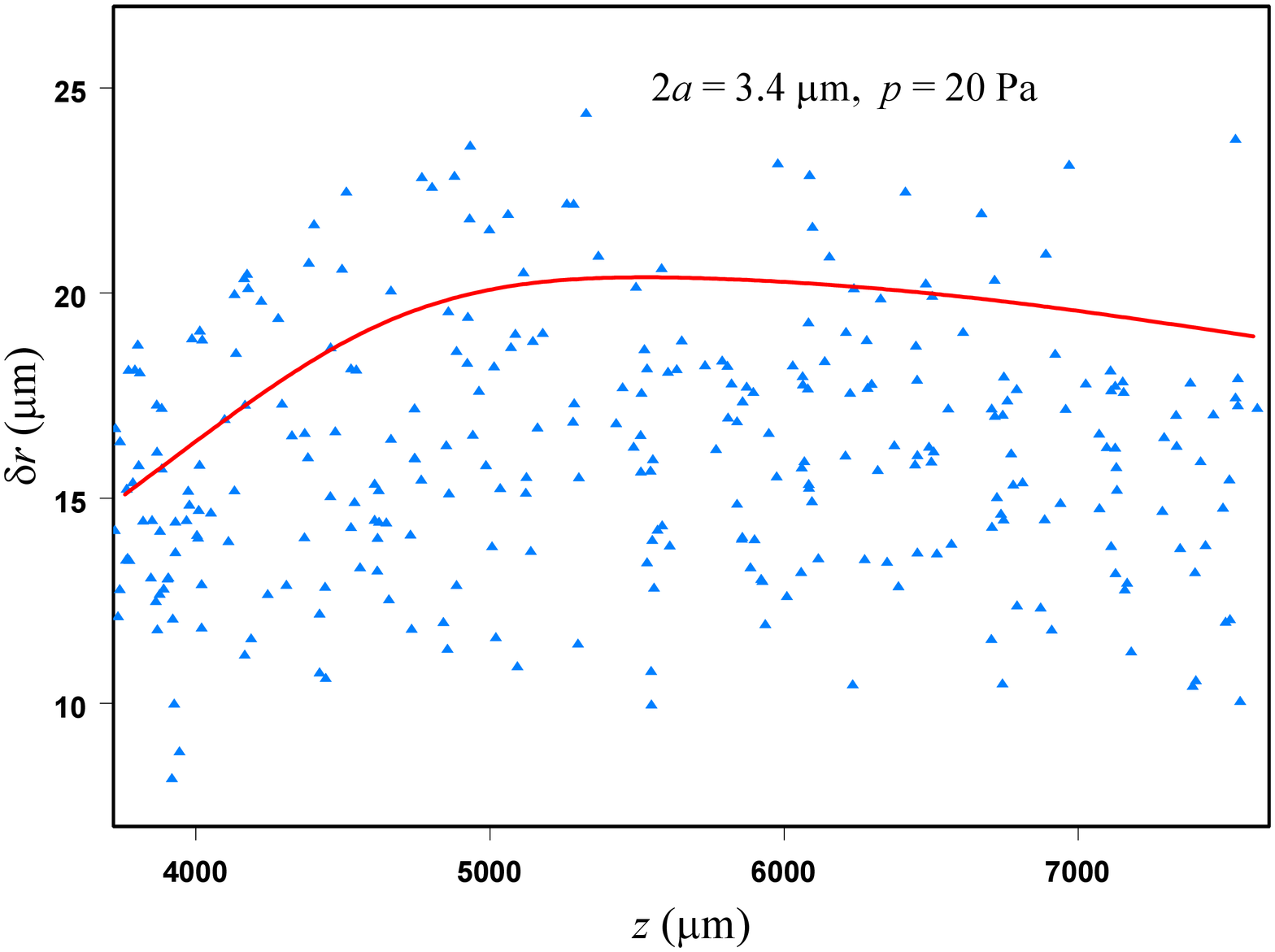}}
\end{picture}
\caption{\label{f10} Same as Fig.~\ref{f8}. The particle 
diameter is $3.4\;\mu {\mbox{m}}$
 and the argon pressure is $20.5\;{\mbox{Pa}}$.}
\end{figure}

\begin{figure}
\centering \unitlength=0.24pt
\begin{picture}(800,770)
\put(-200,-20){\includegraphics[width=10.6cm]{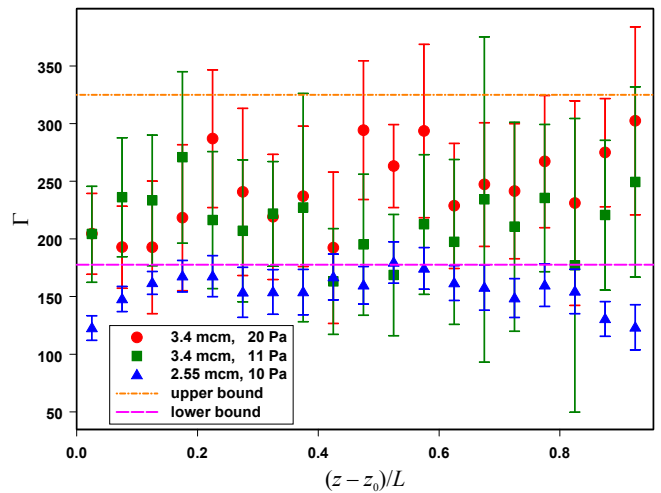}}
\end{picture}
\caption{\label{f11} Coupling parameter for the subsystem 
of dust particles vs the relative distance from the void 
boundary along the discharge axis ($z_0$ is equal to the 
minimum $z$
 for corresponding dust cloud; see 
Figs.~\ref{f8}--\ref{f10}). Dots represent the results of data 
processing for different experiments: circles, for $2a = 
3.4\;\mu {\mbox{m}}$
 and $p = 20.5\;{\mbox{Pa}}$; squares, for $2a = 3.4\;\mu 
{\mbox{m}}$
 and $p = 11\;{\mbox{Pa}}$; and triangles, for $2a = 
2.55\;\mu {\mbox{m}}$
 and $p = 10\;{\mbox{Pa}}$. Dashed and dash-and-dot line 
indicate the lower and upper bounds from the theory 
(formula (\ref{e4})), respectively, for the above-mentioned 
experimental conditions.}
\end{figure}

We used the video frames recorded in experiment 
Ref.~\cite{69} and the particle number density distributions 
obtained in this study to calculate the radius-vector standard 
deviation for the dust particles $\delta r$
 and the coupling parameter for the dust subsystem 
$\Gamma$ (formulas (\ref{e031}) and (\ref{e4}), 
respectively) for three sets of the particle radius and argon 
pressure. The results are summarized in Table~\ref{t1} and 
in Figs.~\ref{f8}--\ref{f11}. As it follows from 
Table~\ref{t1}, $\delta r$
 increases both with the increase in the particle diameter 
and the gas pressure and these dependences are rather 
weak. At first sight, the increase of $\delta r$
 with the gas pressure seems to contradict the increase in 
the friction coefficient $\nu $. However, the behavior of 
$\delta r$
 is also a result of the change in plasma parameters as the 
pressure is changed. In particular, the dependences $n_d 
(z)$
 for different sets are pressure-dependent. It is interesting to 
note that the ratio $\delta r/r_d$ for the sets 1--3 is almost 
constant and it ranges from 0.13 to 0.11. The plots of 
$\delta r$
 vs $z$
 that were obtained by processing the particle traces 
(``clews'') using formula (\ref{e031}) are shown in 
Figs.~\ref{f8}--\ref{f10}. In these figures, the regions near 
the void boundary and the near-electrode sheath are not 
shown because in the former, the particles are unstable and 
in the latter, there is no dust crystal. In spite of a significant 
dispersion of data points, the trend is obvious. In 
Figs.~\ref{f8} and \ref{f9}, $\delta r$
 increases with $z$
 and in Fig.~\ref{f10}, $\delta r$
 has a maximum. It can be easily seen that all these 
dependencies just follow corresponding dependencies $r_d 
(z)$
 (cf.\ Figs.~5--7, Ref.~\cite{69}). The fact that $\delta 
r/r_d$ is independent of the coordinates follows from 
Eq.~(\ref{e21}). One can see that theoretical estimates 
agree satisfactorily with the results of experimental data 
processing.
\begin{table*}
\caption{\label{t1}Radius-vector standard deviation of the 
dust particles $\delta r$
 and the coupling parameter of dust subsystem $\Gamma$ 
in the center of a dust crystal for the experiments with 
different particle diameter $2a$, argon pressure $p$, and 
the Wigner-Seitz cell radius $r_d$ estimated from 
experimental data of Ref.~\cite{69}. For the relative 
standard deviation of the particle charge $\sigma _Z$ and 
the kinetic dust temperature $T_d $, theoretical estimations 
for the conditions of corresponding experiments are 
presented.}
\begin{ruledtabular}
\begin{tabular}{cccccccc}
Set \# & $2a\;{\mbox{(}}\mu {\mbox{m)}}$
 & $p\;{\mbox{(Pa)}}$
 & $r_d \;{\mbox{(}}\mu {\mbox{m)}}$
 & $\delta r\;{\mbox{(}}\mu {\mbox{m)}}$
 & $\Gamma$ & $\sigma _Z$ & $T_d \;{\mbox{(eV)}}$
 \\ \hline
1 & 2.55 & 10 & 90 & 12 & 160 & 0.144 & 1.2 \\
2 & 3.4  & 11 & 136 &16 & 220 & 0.157 & 3.1 \\
3 & 3.4 & 20.5 & 157 & 17 & 250 & 0.195 & 4.2 \\
\end{tabular}
\end{ruledtabular}
\end{table*}

For the sets 1--3, the coupling parameter ranges from 160 to 
250 (Table~\ref{t1}, Fig.~\ref{f11}). This result can be 
compared with the crystallization threshold for OCP 
$\Gamma = 168 \pm 4$
 \cite{88,98}. Apparently, the dust crystal finds itself far 
from the solid--liquid binodal, so that for this system, the 
coupling parameter at binodal must be noticeably lower 
than for OCP. However, one can testify that $\Gamma \sim 
100$
 for both systems. This means that the dust crystal is an 
analog of OCP but these systems are not identical. Note that 
the same order-of-magnitude estimation ($\Gamma \sim 
100$) was reported for the Yukawa ball in the experimental 
study \cite{97}. Figure~\ref{f11} illustrates the absence of 
a noticeable coordinate dependence of the coupling 
parameter. In Fig.~\ref{f11}, the data points from 
individual ``clews'' are averaged within the intervals of $z$
 equal to the difference between successive values of the 
coordinate. The fact that $\Gamma$ is independent of the 
coordinate follows from formula (\ref{e22}). One can see 
that the results of experiment processing lie within the 
minimum and maximum $\Gamma $'s (\ref{e22}) 
calculated for the sets 1-3, i.e., the theory proposed in 
Sec.~\ref{s4} agrees satisfactory with the experiment. Note 
that this theory does not match correctly the pressure 
dependence of $\Gamma $. This could be a consequence of 
(a) simplifying assumptions made, (b) inaccuracy of the 
IEOS (\ref{e11}) and (\ref{e12}), and (c) a relatively high 
sensitivity of $\Gamma$ to $p$.

Knowledge of $\Gamma$ allows one to estimate the 
particle kinetic temperature $T_d = Z_0^2 e^2 /r_d 
\Gamma$ (Table~\ref{t1}). Surprisingly, $T_d$ amounts to 
several eV, which is anomalously high and exceeds the gas 
temperature $T_n$ by two orders of magnitude. It is 
noteworthy that the particles in complex plasmas are 
somewhat overheated due to the ion fluxes to their surface. 
However, the temperature of particle material cannot 
exceed ca.\ $450\;{\mbox{K}}$
 because this is the temperature of evaporation. Some 
degradation of particles during the experiments was in fact 
revealed \cite{93}. Thus, high $T_d$ is solely of the kinetic 
nature, so we can use the term ``anomalous kinetic 
heating.'' It is important that for the first time, this 
phenomenon was observed for a stationary 3D dust crystal. 
In contrast, measurement of the particle velocities in 2D 
dust crystals results in the particle kinetic temperature close 
to room temperature; a substantial increase of this 
temperature was observed only for the fluid and gaslike 
states \cite{3,94,95}.

The above estimate for the particle kinetic temperature 
includes the particle charge $Z_0 $, which was estimated 
using the OML approximation. It was demonstrated that the 
effect of the ion-neutral collisions \cite{67,60} decreases 
the calculated particle charge. However, it follows from 
these studies that this effect is small at the gas pressure less 
than $30\;{\mbox{Pa}}$
 and $Z_0 n_d /n_e > 1$. Such conditions are typical for the 
experiments treated in this work. Thus, the particle kinetic 
temperatures listed in Table~\ref{t1} seem to be somewhat 
overestimated. We recall that the Coulomb coupling 
parameter $\Gamma$ (\ref{e4}) needs no correction for this 
effect because it is independent of $Z_0 $.

The effect of anomalous kinetic heating is a result of 
relatively high particle charge fluctuations. According to 
the discussion in Sec.~\ref{s4} the charge fluctuations for a 
particle in a dust crystal is much larger than that for a 
solitary particle. As indicated in Table~\ref{t1}, the relative 
standard deviation (\ref{e20}) $\sigma _Z \sim 0.1$
 while for a solitary particle it would be an order of 
magnitude lower, $\sigma _Z = 1/\sqrt {(1 + \Phi _0 )Z_0 } 
\sim 0.01$.

\section{\label{s6} CONCLUSION}

To summarize, we have developed a method of the dust 
coupling parameter determination that utilizes sequences of 
video frames recorded in experiments performed on the 
PK-3 Plus setup. We have demonstrated that the particle 
coupling parameter $\Gamma$ is defined by solely the 
standard deviation of the particle radius-vectors and the 
local particle number density. Thus, there is no need for the 
information on the particle charge and velocity, the 
oscillation frequency, etc. For the particle number density, 
we borrowed the distributions determined in our previous 
study \cite{69}. The peculiarities of analyzed oscillations of 
particles in their Wigner--Seitz cells are as follows. The 
oscillations prove to be purely isotropic (within the 
experimental accuracy) in the entire volume occupied by 
the dust crystal, which is indicative of the fact that the 
particle oscillations are not related to the ambipolar electric 
field and the ion flux in complex plasma. Within the 
investigated volume, the oscillations are almost 
homogeneous along the $X$-axis and weakly 
inhomogeneous in the direction of the $Z$-axis. This 
mimics exactly the coordinate dependence of the particle 
number density. The Gaussian form of the probability of 
particle shift from the center of its cell is indicative of the 
fact that the particles oscillate in the spherically symmetric 
quadratic potential. The oscillations of neighboring 
particles are correlated. Their kinetic temperature is 
anomalously high and exceeds the gas temperature by two 
orders of magnitude.

The theoretical interpretation of this anomalous heating 
implies the effect of particle charge fluctuations. In a dust 
crystal, the local particle charge and the electron and ion 
number densities are self-consistent variables. Based on the 
IEOS and the equation for the perturbation evolution in 
complex plasma, we have demonstrated that in this case, 
the amplitude of charge fluctuations is much greater than 
for a solitary particle in infinite plasma even in the absence 
of an instability. This amplitude is sufficiently high to 
ensure a significant effect on the particle oscillation 
amplitude. Proposed interpretation is rather qualitative 
because it utilizes a number of crude assumptions like the 
applicability of the fluid approach at the length scale of the 
interparticle distance. However, the derived formulas allow 
one to make order-of-magnitude estimates for the particle 
standard deviations. Apparently, a more rigorous theory 
would require the treatment of a self-generated phonon 
background in the dust crystal. Development of such theory 
will be addressed in the future.

The theoretical estimates agree satisfactorily with the 
results of processing the data of experiment \cite{69} and 
point to the anomalous kinetic heating of particles under 
stationary conditions. Both the theory and the experiment 
lead to the particle coupling parameter $\Gamma \sim 100$, 
which is of the same order as that for OCP at the binodal of 
the solid--liquid phase transition. This allows us to account 
qualitatively for the observed kinetics of phase transition in 
complex plasma. This investigation can be a basis for the 
development of a theory of phase transitions in strongly 
coupled complex plasmas.

\begin{acknowledgments}
This research is supported by the Russian Science 
Foundation, Grant No.~14-12-01235.
\end{acknowledgments}
\providecommand{\noopsort}[1]{}\providecommand{\singleletter}[1]{#1}%
\end{document}